\documentclass[aps, prx, reprint, showpacs, superscriptaddress,longbibliography]{revtex4-1}

\usepackage{graphicx}
\usepackage{dcolumn}
\usepackage{bm}
\usepackage{hyperref}
\usepackage{amsmath}

\newcommand{\singletS}{${}^1$S$_0\,$}
\newcommand{\singletP}{${}^1$P$_1\,$}

\newcommand{\tripletPj}{$^3$P$_J\,$}

\newcommand{\tripletPone}{$^3$P$_1\,$}
\newcommand{\tripletPzero}{$^3$P$_0\,$}
\newcommand{\tripletPtwo}{$^3$P$_2\,$}

\usepackage{physics}
\usepackage{float}
\usepackage{xcolor}

\begin{document}

\title{Enhancing optical lattice clock coherence times with erasure conversion}

\author{Shuo Ma$^{\dagger}$}
\author{Jonathan Dolde$^{\dagger}$}
\author{Xin Zheng$^{\dagger}$}
\author{Dhruva Ganapathy}
\author{Alexander Shtov}
\author{Jenny Chen}
\author{Anke St{\"o}ltzel}
\author{Bennett J. Christensen}
\author{Shimon Kolkowitz}
\email{To whom correspondence should be addressed; \\
Email: kolkowitz@berkeley.edu}

\affiliation{
 Department of Physics, University of California, Berkeley, CA 94720, USA
}

\date{\today}

\begin{abstract}
Increasing coherent interrogation times is central to advancing the precision of optical clocks.
Synchronous differential optical clock comparisons have now demonstrated atomic coherence times that far exceed the coherence time of the clock laser. While atom coherence times are then primarily limited by errors induced by lattice Raman scattering, excited clock state radiative decay, and broadening from two-body collisions, many of these errors take the atoms out of the clock transition subspace, and can therefore be converted into ``erasure'' errors if the appropriate readout scheme is employed. Here we experimentally demonstrate a hyperfine-resolved readout technique for ${}^{87}$Sr optical lattice clocks that mitigates decoherence from Raman scattering induced by the lattice as well as radiative decay.
By employing hyperfine-resolved readout in synchronous differential comparisons between  ${}^{87}$Sr ensembles with both Ramsey and spin echo spectroscopy sequences, we achieve enhanced atomic coherence times exceeding 100 s and 150 s, respectively, enabling longer coherent measurements without a reduction in performance.
We anticipate that this hyperfine-resolved readout technique will benefit applications of state-of-the-art optical lattice clock comparisons
in which the coherence times are constrained by Raman scattering or radiative decay.

\end{abstract}

\maketitle

\section{Introduction}

State-of-the-art optical lattice clock comparisons have reached measurement precision at the $10^{-20}$ level~\cite{bothwell_resolving_2022,zheng_differential_2022} and absolute clock accuracy at or below the $10^{-18}$ level~\cite{SingleIon_-18_2016,mcgrew_atomic_2018,bothwell_jila_2019,Brewer_IonClock_2019,aeppli_clock_2024}.
This enables a range of novel emerging applications, such as relativistic geodesy, searches for ultralight dark matter, and gravitational wave detection~\cite{
chou_optical_2010,takano_geopotential_2016,takamoto_test_2020,zheng_redshift_2023,derevianko_hunting_2014,kennedy_dark_2020,kolkowitz_gravitational_2016,safronova_search_2018,ludlow_optical_2015}.
Central to these applications are low clock comparison instabilities and interrogation times that fully exploit the long lifetimes of the clock states.

While the performance of an optical lattice clock as an absolute time-keeper or frequency reference is limited by the coherence of the optical local oscillator~\cite{Oelker_ClockStability_2019, ludlow_optical_2015}, synchronous differential comparisons between clocks sharing the same laser light can achieve coherence times exceeding the atom-laser coherence and can extend the coherent clock interrogation time by orders of magnitude.
Recent demonstrations with neutral Sr atoms include differential comparisons between spatially separated atom ensembles in a one-dimensional (1D) optical lattice ~\cite{zheng_differential_2022}, and
subregions of an extended atom ensemble in a cavity-enhanced 1D optical lattice clock~\cite{bothwell_resolving_2022}.
However, the demonstrated atomic coherence times have not yet reached the expected excited clock state lifetime in $^{87}$Sr,
which is believed to be on the order of $120-160~$s~\cite{ross_prl_2019,Thompson_prr_2021,Lu2024,Jack_lifetime_2025}, with one measurement even reporting a lifetime of 330(140)s \cite{ross_prl_2019,Thompson_prr_2021,Lu2024,Jack_lifetime_2025,Dorscher_pra_2018}, with Raman scattering induced by trapping light as the primary limiting factor~\cite{zheng_differential_2022,Dorscher_pra_2018,ross_prl_2019,Jack_lifetime_2025}.   

Standard measurement protocols employed in optical lattice clocks do not distinguish between coherent atoms and those affected by errors such as Raman scattering. Here we show that the decoherence caused by Raman scattering and radiative decay in optical lattice clocks can be partially mitigated by identifying and removing atoms that have scattered photons or radiatively decayed, significantly extending the achievable coherent interrogation times. This concept is inspired by erasure conversion in quantum error detection~\cite{Wu_NC_2022,Ma_Nature_2023,Endres_Nature_2023,Levine_PRX_2024}, and a related strategy was recently theoretically proposed to mitigate errors and enhance the performance of optical clocks and quantum sensors~\cite{Pradeep_PRL_2024}. 

In this work, erasure conversion is realized for ${}^{87}$Sr atoms in a multi-ensemble optical lattice clock by directly probing the coherent clock population using a hyperfine-state-resolved readout method, discriminating between the atoms that remain in the clock subspace and those that have scattered into other hyperfine states. This method introduces minimal experimental overhead, requiring only three additional $\pi$-pulses on the narrow-linewidth clock transition. As these pulses are applied after clock interrogation and only change the method of state readout, we do not expect this technique will introduce additional differential systematic frequency shifts.\par

Using hyperfine-resolved erasure conversion we roughly double the achievable atomic coherence times in our system, observing atomic coherence times exceeding 100 s with synchronous Ramsey spectroscopy and over 150 s with a  synchronous spin echo sequence. We introduce a theoretical framework and perform numerical simulations incorporating known decoherence sources, including Raman scattering, lattice-induced AC Stark shifts, and density shifts, to explain the origin of this coherence enhancement. While we do not yet realize a meaningful enhancement in the measured stability of the clock comparison compared to standard readout, we find that erasure conversion flattens out the scaling of clock instability with interrogation time, offering advantages in applications requiring flexible interrogation times, such as gravitational wave detection\cite{kolkowitz_gravitational_2016}. Additionally, extending coherence times with erasure conversion could also enable new applications, such as spectroscopy of splittings below the frequency resolution limit set by the natural linewidth\cite{OBRIEN19851,gefen2019overcoming}.

\section{A multi-ensemble optical lattice clock}

\begin{figure}[tbp]
\includegraphics[]{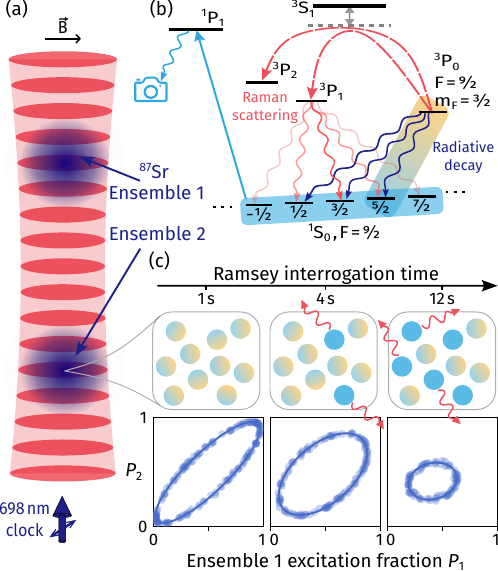}
\caption{\label{Fig1} Lattice photon-induced Raman scattering and its effect on hyperfine state populations. (a) Two ensembles of $^{87}$Sr atoms, initially prepared in the clock subspace, are simultaneously probed by the 698 nm clock laser under a magnetic bias field. (b) The relevant energy level diagram for a $^{87}$Sr optical lattice clock. Atoms are initialized in the clock subspace, consisting of $\ket{^1\text{S}_0, m_F=5/2}$ (blue) and $\ket{^3\text{P}_0, m_F=3/2}$ (yellow). Raman scattering from the lattice light redistributes the population from $\ket{^3\text{P}_0, m_F=3/2}$ to the \tripletPj manifolds, while radiative decay directly returns excited-state atoms to \singletS. Radiative decay from \tripletPone further redistributes the population among the \singletS $m_F$ states. Detection is performed via fluorescence imaging on the \singletS-\singletP transition. (c) The excitation fractions of ensembles 1 and 2 from synchronous Ramsey measurement are plotted parametrically, resulting in ellipses determined by the relative detuning between the clock transitions of the atoms in the two ensembles. Over time, Raman scattering and radiative decay transfer atoms out of the clock subspace, leading to the shrinking size of the ellipses, which corresponds to decoherence.}
\end{figure}

The experimental setup used here is similar to the apparatus described in our previous work~\cite{zheng_redshift_2023}.
As shown in Fig.\ref{Fig1}(a), the experiment begins with the preparation of two ensembles of $^{87}$Sr atoms in a single 1D optical lattice at the magic wavelength near 813.4 nm. We laser cool up to $10^6$ atoms to a temperature of $1~\mu$K using a standard two-stage magneto-optical trapping (MOT) technique. 
Two resolved atom ensembles are prepared by shifting the location of the MOT field zero roughly midway through the single-frequency stage of the narrow linewidth MOT \cite{Hassan_2024}, so that the atoms are cooled into the lattice in two different regions spatially separated from each other by roughly 1 mm.

We perform in-lattice sideband cooling (axially) and Doppler cooling (radially) on the ${}^1\text{S}_0\rightarrow {}^3\text{P}_1 (F=11/2)$ transition. We then spin polarize all the atoms into the stretched state $\ket{^1\text{S}_0,9/2}$, which we label as $\ket{g,9/2}$, by illuminating atoms with circularly polarized 689 nm laser resonant on the ${}^1\text{S}_0\rightarrow {}^3\text{P}_1 (F=9/2)$ transition.
Next, we transfer the atoms from $\ket{g,9/2}$ to $\ket{^3\text{P}_0,3/2}$, which we label as $\ket{e,3/2}$, using three consecutive $\pi$ pulses
on resonance with the $\ket{g,9/2} \rightarrow \ket{e, 7/2}$, the $\ket{e,7/2} \rightarrow \ket{g, 5/2}$, and the $\ket{g,5/2} \rightarrow \ket{e, 3/2}$ transitions, respectively. Each pulse's intensity and frequency are independently calibrated to maximize the transfer efficiency.
During the above sequence, the lattice is held at a depth of $\sim$100 $E_{\rm rec}$ (lattice photon recoil energy).
For clock interrogation, the lattice is then adiabatically ramped down to operational depths, between $\sim$15 and 75 $E_{\rm rec}$.

\section{Impact of Raman Scattering and Radiative decay on Atomic Coherence}

During the synchronous Ramsey comparisons, atoms in both ensembles are prepared in a superposition of $\ket{g, 5/2}$ and $\ket{e, 3/2}$, as shown in Fig.\ref{Fig1}(b), the least magnetically sensitive clock transition for $^{87}$Sr, which we define as the clock subspace. Because a clock interrogation sequence ends with a projective measurement of the remaining populations in $\ket{g, 5/2}$ and $\ket{e, 3/2}$, atoms lost during interrogation do not contribute signal or noise to the final measurement outcome. 
However, Raman scattering from the lattice trapping light redistributes population from \tripletPzero into \tripletPone and \tripletPtwo, with calculated branching ratios of 0.39 and 0.22, respectively. Meanwhile, 0.39 of the population undergoes Rayleigh scattering, remaining in \tripletPzero without flipping the nuclear spin state, due to the large detuning relative to the hyperfine splitting in the excited state and the use of linearly polarized lattice light (see Appendix \ref{Appendix: Raman scattering} for details). In particular, Raman scattering from \tripletPzero to \tripletPone, followed by subsequent radiative decay, redistributes population into the \singletS manifold with $\Delta m_F = 0, \pm1, \pm2$ according to selection rules, as depicted in Fig.\ref{Fig1}(b)\&(c). We find that most atoms scattered into \tripletPtwo subsequently escape the trap due to high two-body collision rates, effectively being erased. However, a small residual \tripletPtwo population would contribute to the final signal after repumping. Rayleigh scattering within \tripletPzero and \singletS does not induce decoherence due to the use of a magic wavelength lattice \cite{Uys2010}, ensuring equal polarizabilities for the $g$ and $e$ states. On the other hand, radiative decay from the excited state redistributes the population among the $m_F$ states in the $^1\mathrm{S}_0$ manifold, with changes of $\Delta m_F = 0, \pm 1$.\par

Standard readout via the \singletS-\singletP transition accounts for the remaining ground and excited state populations but excludes atoms lost during interrogation. However, it cannot resolve contributions from different $m_F$ populations, leading to coherence degradation in projective measurements, as shown in Fig.~\ref{Fig1}c. In contrast, hyperfine-resolved readout selectively probes the clock subspace, distinguishing coherent atoms and thereby preserving and extending coherence.

\section{Measurement of hyperfine state population distribution induced by Raman scattering}

\begin{figure}[tp]
\includegraphics[]{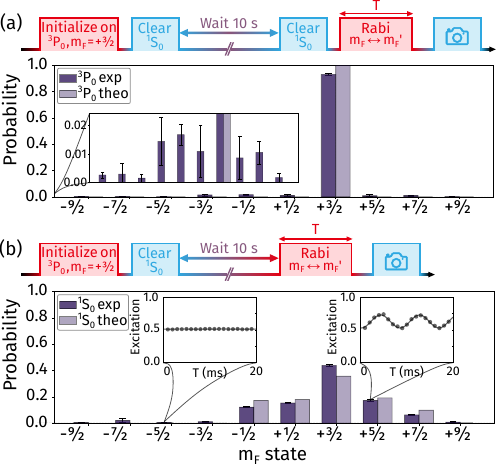}
\caption{\label{Fig2} Hyperfine state population ratio measurement of erasures. The top panel of (a) illustrates the sequence used to measure the excited-state population distribution. The system is first initialized in $\ket{e, 3/2}$, followed by a clean-up pulse to remove the ground-state population. After holding the atoms in the lattice for 10 s, another clean-up pulse eliminates remaining ground-state atoms. A specific clock transition $\ket{g, m_F} \leftrightarrow \ket{e, m_F'}$ is then driven for varying pulse durations to map out the $\ket{g, m_F}$ state population. The bottom panel displays the resulting excited-state population distribution, with an inset that zooms into the probability below 0.02 in order to show the population in $m_F$-states other than $\ket{e, 3/2}$ more clearly. Similarly, the top panel of (b) depicts the sequence used to measure the ground-state population. The bottom panel presents the corresponding population distribution. Dark purple bars indicate measured normalized $m_F$-state populations in $g$ (lower panel) and $e$ (upper panel), while light purple bars indicate the theoretically predicted populations. Insets: Representative Rabi oscillation plots of the $\ket{g,-5/2}\leftrightarrow\ket{e,-7/2}$ transition (left panel) and the $\ket{g,5/2}\leftrightarrow\ket{e,7/2}$ transition (right panel), respectively.}
\end{figure}

In order to determine the erasure conversion efficiency of hyperfine-resolved readout, we experimentally measure the nuclear spin population distribution for spin-polarized atoms prepared in the excited clock state and subject to Raman scattering and radiative decay at a trap depth of 100 $E_{\text{rec}}$. The experimental sequence is shown at the top of Fig.~\ref{Fig2}(b). We first initialize atoms in $\ket{e, 3/2}$ and then apply a clean-up pulse on the ${}^1\text{S}_0 \rightarrow {}^1\text{P}_1$ transition to remove residual population in $g$ caused by pulse imperfections. Atoms are held in the lattice at a trap depth of approximately $100~E_{\rm rec}$ for 10 s. As the radiative decay and Raman scattering rates have different dependencies on trap depth, the erasure fraction varies accordingly. We select this depth to ensure a sufficient number of scatterings and decays returning to the ${}^1\text{S}_0$ state while maintaining a reasonable overall measurement cycle time.

We spin-selectively probe each of the $\ket{g, m_F}$ population using hyperfine-state-dependent clock transitions ($\ket{g, m_F} \leftrightarrow \ket{e, m_F^{\prime}}$). We calibrate the transition frequency for each hyperfine clock transition and perform Rabi oscillations by selectively driving one clock transition at a time. We average the result for each transition over at least 10 measurements and fit to a sinusoid. We infer the population of the $m_F$ state from the fitted amplitude of the Rabi oscillation. Examples of the measured and fitted Rabi oscillations for the $\ket{g, 5/2} \leftrightarrow \ket{e, 7/2}$ and $\ket{g, -5/2} \leftrightarrow \ket{e, -7/2}$ transitions are shown in the inset of Fig.\ref{Fig2}(b). A substantial population is observed in $\ket{g, 5/2}$, 
while little population is detected in $\ket{g, -5/2}$.\par

Summing the fitted amplitudes for all 10 ${}^1\text{S}_0$ hyperfine state populations,
we obtain a total population of $0.449(19)$. 
Adding this to the average remaining \tripletPzero and \tripletPtwo population of $0.541(1)$ results in a total amplitude of $0.990(20)$, 
consistent with unity.
For clarity, in Fig.~\ref{Fig2}(b)
we present the normalized population distribution of \singletS, obtained after subtracting the remaining \tripletPzero and \tripletPtwo populations. We compare this to the theoretically calculated population distribution and find a reasonable agreement between theory and experiment. The calculated Raman and Rayleigh scattering rates are tabulated in Appendix \ref{Appendix: Raman scattering}. We attribute the small but statistically significant deviations between the experiment and the theoretical prediction mainly to scattering contributions from higher excited states and from the omission of hyperfine splitting in the excited states in our calculations.

We measure the population distribution of \tripletPzero using a sequence nearly identical to that in Fig.\ref{Fig2}(b), with the addition of a clean-up pulse to remove \singletS populations before driving the clock pulse, as shown at the top of Fig.\ref{Fig2}(a).
The sum of the ten fitted Rabi oscillation amplitudes is 0.919(19).
We suspect that the deviation from unity arises from residual population in the \tripletPtwo state due to scattering.
Our measured probability distribution of the $e$ hyperfine states (after normalization) indicates that the majority of the population undergoes $\Delta m_F = 0$ transitions. However, a small fraction appears in other excited $m_F$ states.
We suspect this arises from the Raman scattering between different excited $m_F$ states caused by the impurity in the lattice laser's polarization.

\begin{figure*}[tp]
\includegraphics[]{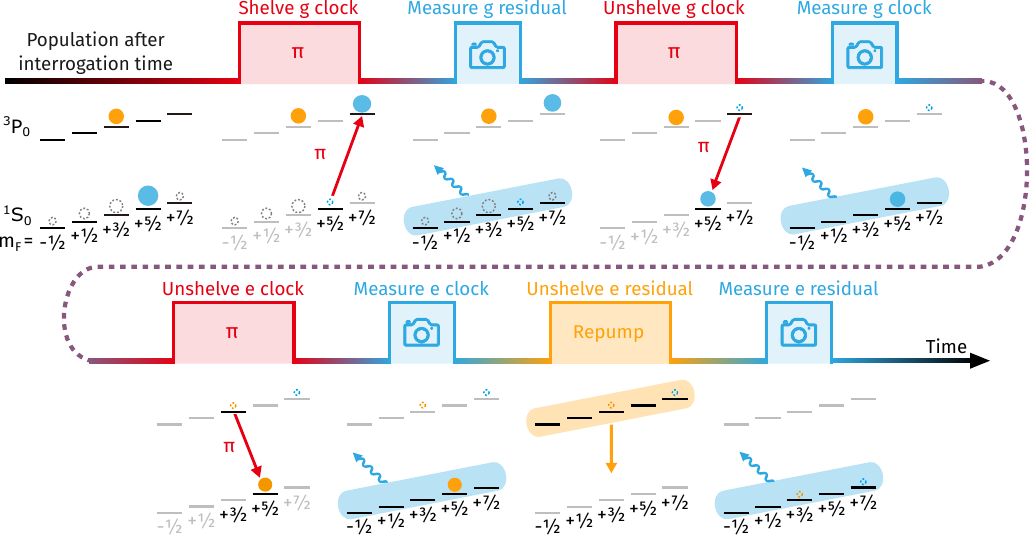}
\caption{\label{Fig3} Illustration of pulse sequence used for hyperfine state resolved readout, starting with a quantum superposition between $\ket{g,5/2}$ and $\ket{e,3/2}$.
The $\ket{g,5/2}$ population is shelved in the clock state by a first $\pi$-pulse on resonant with the $\ket{g,5/2} \rightarrow \ket{e,7/2}$ transition; The remaining populations in $g$ is measured ($N_{g,res}$) using a global imaging pulse ($\sim25~\mu$s) on the ${}^1\text{S}_0 \rightarrow {}^1\text{P}_1$ transition and at the same time erased from the remaining operations;
A second clock $\pi$-pulse resonant with the $\ket{e,7/2} \rightarrow \ket{g,5/2}$ transition is used to coherently unshelve the population back to the ground state;
The $\ket{g,5/2}$ population ($N_{g,5/2}$) is measured using a second imaging pulse;
The $\ket{e,3/2}$ population is transferred to $g$ via a third clock pulse on the $\ket{e,3/2} \rightarrow \ket{g,5/2}$ transition;
The $\ket{e,3/2}$ population ($N_{e,3/2}$) is measured using a third imaging pulse;
The remaining populations in ${}^3\text{P}_0$ and ${}^3\text{P}_2$ are transferred back to $g$ using standard global repump pulses on the ${}^3\text{P}_0\rightarrow{}^3\text{S}_1$ and  ${}^3\text{P}_2\rightarrow{}^3\text{S}_1$ transitions;
A final imaging pulse is applied to measure the residual excited state population ($N_{e,res}$).
}
\end{figure*}

A comparison between the experimentally measured and theoretically calculated populations in $g$ (after normalization) is shown in Fig.~\ref{Fig2}(b). We observe that 
roughly 18\% of atoms that are subject to Raman scattering are found in the $\ket{g, 5/2}$ state. This sets a lower bound on the detectable erasures through hyperfine-resolved readout. However, this limitation could in principle be mitigated by using more widely separated hyperfine clock states, such as a superposition between $\ket{g, 9/2}$ and $\ket{e, 3/2}$. This approach would come at the cost of increased magnetic field sensitivity and would require additional clock or microwave pulses for coherent population transfer.

\section{Hyperfine-state-resolved readout method}

In Fig.~\ref{Fig3}, we present a protocol for erasure conversion via hyperfine-resolved readout, which utilizes a sequence of clock $\pi$-pulses to coherently transfer and selectively measure population in specific targeted nuclear spin states. The protocol begins by shelving the $\ket{g, 5/2}$ population using a clock $\pi$-pulse resonant with the $\ket{g, 5/2} \rightarrow \ket{e, 7/2}$ transition. An imaging pulse then measures the remaining residual ground-state population, $N_{g, \text{res}}$, while also removing them. We then apply a second clock $\pi$-pulse on the same transition, followed by another imaging pulse to measure the shelved $\ket{g, 5/2}$ population, $N_{g, 5/2}$.
Next, a third clock $\pi$-pulse, resonant with the $\ket{e, 3/2} \rightarrow \ket{g, 5/2}$ transition, transfers the $\ket{e, 3/2}$ population back to $g$, where it is measured with a third imaging pulse. A global repump pulse on the ${}^3\text{P}_{0,2} \rightarrow {}^3\text{S}_1$ transitions, followed by a fourth imaging pulse, measures the residual excited-state population, $N_{e, \text{res}}$. Finally, a background imaging pulse without atoms records the stray light background, $N_{bg}$. The entire sequence is completed in under 50 ms.

The excitation fraction in the clock subspace probed by hyperfine-resolved readout is calculated using:
\begin{equation}
    P_{hf} = \frac{N_{e, 3/2} - N_{bg}}{N_{e, 3/2} + N_{g, 5/2} - 2 N_{bg}},
\end{equation}
while the reconstructed excitation fraction for standard readout, using the same data but including the measured residual atoms, is expressed as:
\begin{equation}
    P_{recon} = \frac{N_{e, 3/2} + N_{e, \text{res}} - 2N_{bg}}{N_{e, 3/2} + N_{g, 5/2} + N_{g, \text{res}} + N_{e, \text{res}} - 4 N_{bg}}.
\label{reconstruct}
\end{equation}

\section{Enhanced atomic coherence}
We find that the atomic coherence is significantly improved by using hyperfine-resolved readout in synchronous differential comparisons employing both Ramsey and spin echo spectroscopy. 
In order to perform the comparisons, two spatially separated ensembles of ${}^{87}$Sr atoms are co-trapped in a single 1D optical lattice and probed simultaneously by a shared clock laser along the axial direction [Fig.~\ref{Fig1}(a)]. 
The excitation fractions of the two ensembles are parametrically plotted and fitted to a resulting ellipse, from which the frequency difference between the two ensembles can be determined, as well as the Ramsey contrast of each atomic ensemble. 
This synchronous measurement bypasses the limitation on interrogation time imposed by the clock laser's phase noise.\cite{zheng_differential_2022}

\subsubsection{Synchronous Ramsey spectroscopy}

We first investigate atomic coherence using synchronous Ramsey spectroscopy\cite{zheng_differential_2022}, where the resulting ellipses are fitted using a maximum likelihood estimator (MLE; see Appendix~\ref{Appendix_A} for details).

\begin{figure*}[tp]
\includegraphics[]{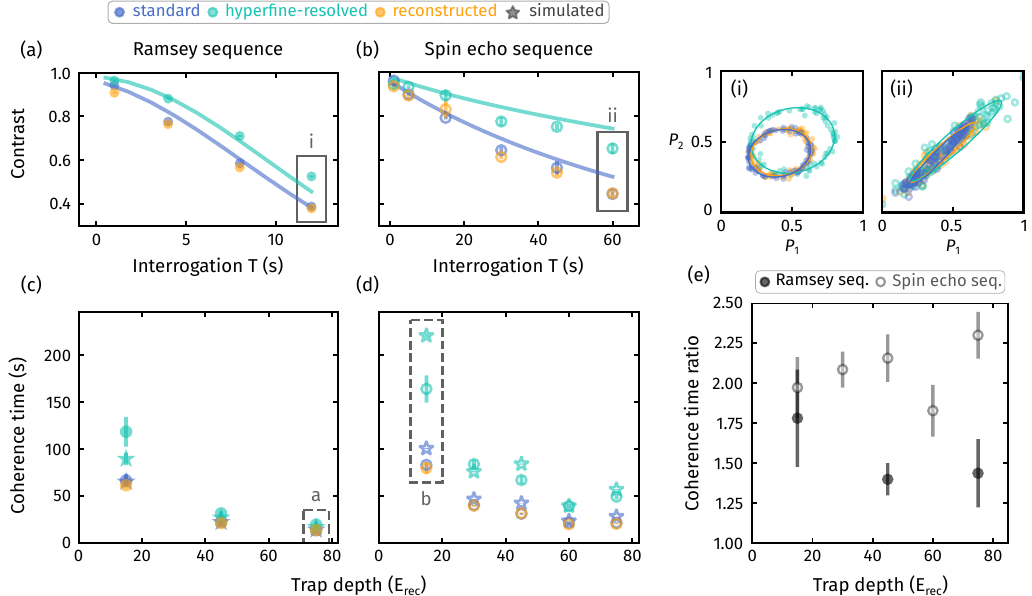}

\caption{\label{Fig_coherence_data} Comparison of coherence times using standard readout (blue), hyperfine-resolved readout (cyan), and reconstructed standard readout (orange) methods. Representative plots of coherence decay are shown in (a) for the Ramsey sequence at 75 $E_{\rm rec}$ (the highest trap depth measured in this work) and in (b) for the spin echo sequence at 15 $E_{\rm rec}$ (the lowest depth), with the corresponding comparison ellipses shown in panels (i) and (ii), respectively. Maximum likelihood estimation (MLE) is used to extract the atomic coherence. The solid curves are calculated through numerical simulation based on the known experimental parameters. (c) shows coherence time versus trap depth for the Ramsey sequence, while (d) presents the coherence time for the spin-echo sequence. Numerically simulated coherence times are indicated by star-shaped markers with matching color coding. (e) compares the enhancement ratio of the measured coherence time using erasure conversion for the Ramsey sequence (solid circles) and spin-echo sequence (hollow circles) as a function of depth. The measured and simulated coherence decay for the other trap depths are presented in Appendix \ref{fig8_no_labels}.}
\end{figure*}

An example of a comparison between the ellipses resulting from standard readout and hyperfine-resolved readout is shown in the inset of Fig. \ref{Fig_coherence_data}(i), 
taken at a trap depth of 75 $E_{\rm rec}$ with a Ramsey free evolution time of 12 s. 
Notably, we observe a significant enhancement in ellipse contrast when using hyperfine-resolved readout (cyan), 
achieving a factor of about 1.5 improvement compared to standard readout. Additionally, the reconstructed ellipse generated from hyperfine-resolved readout (orange) using Eq.~\ref{reconstruct} agrees well with the standard readout ellipse (blue).

In Fig.~\ref{Fig_coherence_data}(a), we show the contrast as a function of coherence time for standard, hyperfine-resolved, and reconstructed readout methods. The fitted $1/e$ coherence time for hyperfine-resolved readout is 19(3)~s, which is a factor of 1.4(1) longer than the coherence time obtained with standard readout (13(1)~s). The numerically simulated contrast versus time using both hyperfine-resolved and standard readout shows reasonable agreement with the data.

We perform synchronous Ramsey spectroscopy at three lattice trap depths: 15, 45, and 75~$E_{\rm rec}$. The fitted $1/e$ coherence times are plotted as a function of trap depth in Fig.\ref{Fig_coherence_data}(c). At 15 $E_{\rm rec}$, we measure coherence times up to 120~s with hyperfine-resolved readout. Additionally, reconstructed standard readout results from hyperfine-resolved readout measurements (cyan filled circles) closely match independently measured standard coherence (blue filled circles).

We numerically simulate the expected coherence times for each trap depth, incorporating the decoherence sources described in Appendix~\ref{Appendix: Numerical simulation}. To model Ramsey spectroscopy, we also account for inhomogeneous dephasing effects such as the magnetic field gradient along the ensemble, measured to be 1.5 mG/cm. Additionally, other inhomogeneous broadening sources, such as residual Stark shifts, whose exact contributions remain uncertain, are modeled through a phenomenological Gaussian dephasing term with an empirically determined rate. 

\subsubsection{Modified synchronous spin echo spectroscopy}

To further reduce decoherence from inhomogeneous broadening, we perform synchronous spin echo spectroscopy under similar experimental conditions. While the standard spin echo sequence effectively cancels much of the low-frequency differential frequency shifts such as those arising from magnetic field and lattice inhomogeneities, it typically also results in a differential phase shift $\phi$ between the two ensembles that is very close to 0 radians.
At this phase, however, 
Least Squares fitting~\cite{ellipse_fitting} is heavily biased due to its numerical constraint enforcing an ellipse-specific solution.
In contrast, maximum likelihood estimation (MLE) intrinsically reduces bias by incorporating a probabilistic noise model, which is particularly beneficial when the ellipse is primarily limited by quantum projection noise (QPN). 
A detailed description of MLE can be found in Appendix~\ref{Appendix_A}.
Although MLE has vanishing bias over a large range of phase, it remains slightly biased when $\phi$ is close to 0 radians because the fit cannot reliably distinguish between positive and negative values of $\phi$, particularly in the presence of QPN. To ensure unbiased MLE, we introduce a transient magnetic field gradient along the lattice axis during the second free precession period of the spin echo sequence (see Appendix~\ref{Appendix_B} for details).
This allows us to fine-tune the differential phase shift by adjusting the magnitude or duration of the applied gradient.
To minimize potential impacts (such as decoherence) from the added transient gradient, we choose the lowest possible gradient ($\approx 2.0$ mG/cm for 1.4 s),
which results in a phase shift of 0.04 radians -- sufficient for MLE to be unbiased.

We perform the modified synchronous spin echo spectroscopy at 5 different lattice trap depths: 15, 30, 45, 60, and 75~$E_{\rm rec}$. 
The $1/e$ coherence times are fitted and plotted as a function of trap depth in Fig.\ref{Fig_coherence_data}(d). 
At a trap depth of 15 $E_{\rm rec}$, we achieve atomic coherence times of up to 150~s after erasure conversion, about twice as long as when using standard readout under the same experimental conditions. 
Again, the reconstructed standard readout from hyperfine-resolved measurements and Eq.~\ref{reconstruct} (cyan empty circles) shows good agreement with independently measured coherence using standard readout (blue empty circles). Our numerical simulations of expected spin echo coherence as a function of trap depth show reasonable agreement with the data.\par

Notably, as shown in Fig.\ref{Fig_coherence_data}(e), the enhancement of coherence time in spin echo spectroscopy with hyperfine-resolved readout surpasses that of Ramsey spectroscopy across all trap depths. In Ramsey spectroscopy, wavefunction collapse induced by the non-Hermitian no-jump operator imposes a finite coherence time \cite{Plenio1998}, even as the erasure fraction approaches 1. In contrast, the population swap due to the $\pi$-pulse in the spin echo sequence effectively erases the origin of the scattering, leading to a significant improvement in coherence time through erasure conversion. A more detailed discussion is provided in Appendix~\ref{Coherence enhancement theory}, and additional plots of all measured coherence decays are included in Appendix~\ref{Appendix: additional plots}.

\section{Resulting differential stabilities and the associated metrological gain}

\begin{figure}[tp]
\includegraphics[]{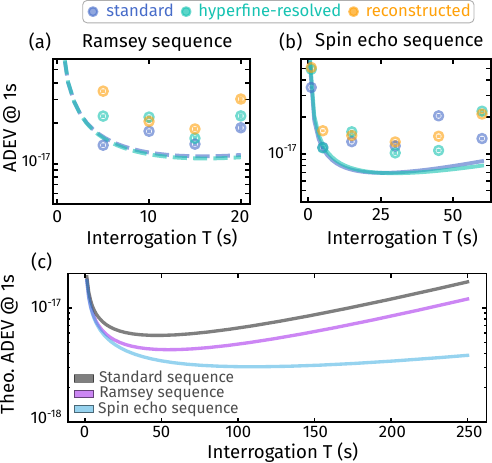}
\caption{\label{Fig_stability} (a)\&(b) show the extracted Allan deviations at 1 second for the differential clock comparisons for Ramsey (filled circles) and spin echo (empty circles), as well as using standard readout (blue), hyperfine-resolved readout (cyan), and reconstructed standard readout (orange), all taken at 15 $E_{\rm rec}$ depth. The error bar of each data point is on the order of $1\times10^{-18}$, and is thus not visible in the plot. The dashed curve in (a) represents the idealized Allan deviation, based only on the simulated coherence and the remaining number of atoms as a function of time, and does not include the relative phase difference between the two ensembles. In (b), the solid lines show the modeled Allan deviations for standard (blue) and hyperfine-resolved (cyan) readout, accounting for the accumulated phase between the two ensembles. Please see Appendix \ref{Appendix: Numerical simulation} for details.} (c) presents the theoretically predicted Allan deviation with 100\% erasure fraction, assuming the only errors are radiative decay and Raman scattering, and 1000 initial number of atoms, for standard readout (black), Ramsey sequence (purple), and spin echo sequence (blue).
\end{figure}

From the same measurements shown in Fig.~\ref{Fig_coherence_data}, we extract differential clock comparison stabilities using the Jackknife resampling technique (Appendix \ref{Appendix_A}). 
Representative Allan deviations (for 1 s averaging time) are presented in Fig.~\ref{Fig_stability} (a)\&(b) for standard and hyperfine-resolved readout, for both synchronous Ramsey and synchronous spin echo sequences. Despite achieving substantial enhancements in atomic coherence times, up to a factor of 1.5 to 2, the extracted stabilities with hyperfine-resolved readout do not show statistically significant improvement over standard readout. This limitation arises because while erasure conversion eliminates the excess noise arising from Raman scattering and radiative decay, the number of atoms that remain coherent and ultimately contribute to the measurement still falls exponentially with interrogation time, leading to a cap on the achievable fundamental improvement of 
$\sqrt{2}$ (or $2$ in the limit of low duty cycle) in optimal clock stability \cite{Pradeep_PRL_2024}. Additionally, the finite erasure fraction and the infidelities (95–98\% per pulse) in the additional clock $\pi$-pulses used for hyperfine-resolved readout further reduce the observed stability gains.
We verify and isolate the impact of the imperfect $\pi$-pulses, which arise due to technical limitations and could in principle be mitigated (see Appendix~\ref{Appendix: Technical limitations},) by comparing the stability achieved with hyperfine-resolved readout to that obtained using the reconstructed readout. Because the reconstructed readout method involves the same number of $\pi$-pulses as the hyperfine-resolved readout, and therefore also suffers from the same infidelities, this comparison reveals an enhancement in the lowest measured differential instability with hyperfine-resolved readout by factors of 1.22(6), 1.08(3), 1.62(6), 1.21(4), and 1.05(3) for lattice depths of 15, 30, 45, 60, and 75~$E_{\rm rec}$, respectively, for the spin echo measurements. We attribute the additional fluctuations in clock stability across readout methods and interrogation times, which significantly exceed the uncertainty in the measured Allan deviation, to variance in the number of initially loaded atoms (detailed in Appendix \ref{Appendix: additional plots}), along with contributions from uncompensated clock laser drift, which impacts the fidelity of the $\pi$-pulses used in the hyperfine-resolved and reconstructed readouts, as well as the relative accumulated phase between the two ensembles, which affects the stability of the differential clocks measurement as detailed in Refs.~\cite{zheng_differential_2022,young_nature_2021, marti_2018} and Appendix~\ref{Appendix: Numerical simulation}.\par

While the benefits of erasure conversion are limited for conventional clock operation or traditional differential comparisons, where the interrogation time can be optimized to minimize the overall instability, it significantly reduces the dependence of instability on the interrogation time. Fig.~\ref{Fig_stability}(c), shows the theoretically calculated clock stability versus interrogation time for standard readout in the Ramsey sequence (black) and hyperfine-resolved readout in the Ramsey (purple) and spin echo (blue) sequences, assuming the contributing error sources including Raman scattering and spontaneous decay at trap depth of 15 $\text{E}_{rec}$ and $100\%$ erasure fraction. The simulation does not include the atom loss from lattice heating. See details about simulation in Appendix \ref{Appendix: Numerical simulation}. The enhanced coherence time in the spin echo sequence results in a flattening out of clock stability compared to standard readout. In applications requiring flexible interrogation times, such as gravitational wave detection \cite{kolkowitz_gravitational_2016} or dark matter searches \cite{Kennedy2020}, this relative insensitivity of stability to interrogation time could provide a significant advantage. Furthermore, the demonstrated extension of coherence time highlights potential applications beyond optical clocks, including the resolution of the line splittings beyond the limit set by the natural linewidth \cite{OBRIEN19851,gefen2019overcoming}.

To further enhance the achievable erasure fraction, one could utilize hyperfine states with larger separations, such as a superposition of $\ket{g, 9/2}$ and $\ket{e, 3/2}$ states which has the lowest magnetic field sensitivity (388 Hz/G) relative to all other transitions with $\Delta m_F\ge$3. This configuration would effectively eliminate erasure contamination from Raman scattering and radiative decay. However, it would come at the cost of increased magnetic field sensitivity compared to the $\ket{g, \pm5/2} \leftrightarrow \ket{e, \pm3/2}$ transition employed here. Implementing this approach would therefore require careful mitigation of stray magnetic field gradients and additional clock $\pi$-pulses for coherent population transfer, imposing stricter requirements on clock pulse fidelities.

Finally, we note that our erasure conversion scheme relies on the large number of nuclear spin states of ${}^{87}$Sr, and is not applicable to optical lattice clocks employing nuclear-spin-less isotopes or isotopes with spin-1/2, such as ${}^{171}$Yb. However, in these systems, as well as in ${}^{87}$Sr, alternative erasure conversion techniques could still be employed, such as the addition of a repumping laser out of ${}^3\text{P}_1$ during clock interrogation to suppress decays to ${}^1\text{S}_0$ and instead transfer scattered atoms into ${}^3\text{P}_2$. Naturally such an approach would come with its own trade-offs, such as large ac Stark shifts of the clock transition, and we leave investigation of this and other alternative approaches to future work. 

\section{Conclusion}

We have experimentally demonstrated an erasure conversion protocol for optical clocks using hyperfine-resolved readout, effectively mitigating decoherence from lattice-light-induced Raman scattering and radiative decay. This approach extends atomic coherence times, achieving Ramsey spectroscopy coherence beyond 100 s and over 150 s with a spin echo sequence. Additionally, we developed a theoretical framework to describe erasure-enhanced coherence in both spectroscopy sequences, supported by numerical simulations incorporating Raman scattering, lattice Stark shifts, and density shifts. Our results demonstrate that hyperfine-resolved erasure detection not only suppresses decoherence from Raman scattering but also has the potential to fully unlock the benefits of ultra-narrow clock transitions, advancing the performance of optical lattice clocks in applications requiring flexible interrogation times. Moreover, coherence enhancement via erasure conversion opens a new avenue for implementing sub-natural linewidth spectroscopy\cite{OBRIEN19851,gefen2019overcoming}.

\begin{acknowledgments}
We thank Jeff Thompson, Jun Ye, Nils Huntemann, Alex Retzker, Alexey Gorshkov, Dimitry Budker, Toby Bothwell, and Andrew Jayich for fruitful discussions and insightful comments on the manuscript. 
This work was supported by a Packard Fellowship for Science and Engineering, the Army Research Office through agreement number W911NF-21-1-0012, the Sloan Foundation, the Simons Foundation, the Gordon and Betty Moore Foundation under grant DOI 10.37807/gbmf12966, NASA under grant No.~80NSSC24K1561, and the National Science Foundation under Grants No.~2143870 and 2326810.

$^{\dagger}$ S.M., J.D., and X.Z.~contributed equally to this work.
\end{acknowledgments}

\section*{Data availability}
The data that support the findings of this article are openly available \cite{paper_data}.

\appendix

\section{Ellipse fitting with maximum likelihood estimator}
\label{Appendix_A}

\begin{figure*}[tp]
\includegraphics[]{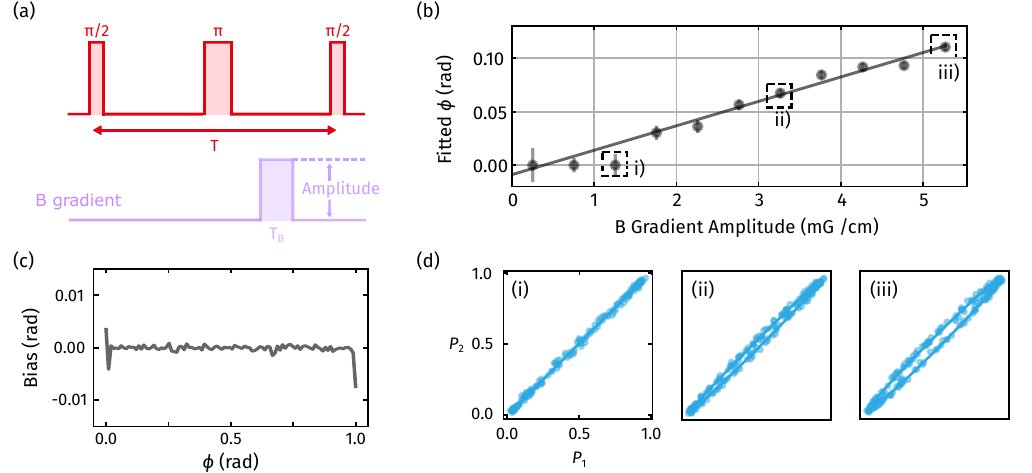}
\caption{\label{Fig_6} Illustration of modified spin echo sequence. (a) Timing sequence
for modified spin echo. To operate away from zero phase shift, a magnetic field (0.5 to 5 mG/cm) is applied to the second free evolution period for time TB =1.4 s.(b) Fitted phase using MLE as a function of B field gradient dB/dz. The solid line is fitted using data points at high gradient ($>$2mG/cm) and shows a fitted offset phase $\phi$ of -0.004(0.007) radians without gradient field, consistent with vanishing bias error using MLE. (c)  Implementation of the MLE algorithm to estimate the relative phase of the ellipse. (d) Callout of specific ellipse fittings using MLE under varying magnetic field gradients.}
\end{figure*}

In order to calculate the differential phase between the two atomic ensembles, we use the angle inferred by fitting the data to an ellipse. For the spin echo sequence, this angle is very close to zero, and the ellipse in close to a line. In this regime, traditional least squares estimators are biased~\cite{ellipse_fitting}. To overcome this limitation, we use Maximum Likelihood Estimation (MLE) \cite{Stockton_2007}. This involves maximizing the likelihood function which quantifies the probability of observing the given data as a function of the model's parameters with the assumption that quantum projection noise is the dominant source of noise in the data. By performing tests on simulated data (see Figure.~\ref{Fig_6}(c)), we see that MLE exhibits no bias over a wide range of phases in ellipse angle estimation. However, bias is not completely eliminated at near-zero differential phases. This is mitigated experimentally by introducing a phase offset, which is discussed in Appendix \ref{Appendix_B}.

The variance of the contrast is determined using a re-sampling technique~\cite{marti_2018} known as jack-knifing :
\begin{equation}
    {\rm var}~C = \frac{N}{N-1}\sum(\bar{C}^{\rm JK} - \bar{C}^{\rm JK}_{\neq i})^2,
\end{equation}
where $N$ is the total number of measurements, and $\bar{C}^{\rm JK}$ is the mean of $\bar{C}^{\rm JK}_{\neq i}$,
which is a fit to all data except the $i^{\rm th}$ data point.

\section{Modified spin echo sequence}
\label{Appendix_B}

While the traditional spin echo sequence effectively removes inhomogeneous broadening at higher Fourier frequencies, it typically results in near-zero differential phase shifts during differential clock comparisons. Despite upgrading our ellipse estimator with MLE, this near-zero phase regime introduces bias. To address this, while still canceling most of the inhomogeneous broadening, we perform a modified spin echo sequence, as depicted in Fig.~\ref{Fig_6}(a).

In this sequence, a fixed-duration, amplitude-tunable magnetic field gradient is applied during the second half of the spin echo period. We fit the ellipses obtained by varying the magnetic field gradient (from 0.02 to 5.5mG/cm) and plot the extracted differential phase shifts as a function of the applied gradient. Several representative fit results are shown in Fig.\ref{Fig_6}(d). Based on these, we choose to operate at a gradient of approximately 2.0~mG/cm, where the phase shift remains linear and the overall phase is kept small, minimizing bias while retaining effective decoherence cancellation.

\section{Raman scattering population distribution calculation}
\label{Appendix: Raman scattering}
We calculate the scattering rate from $\ket{^3\text{P}_0, m_{F}}$ to $\ket{^3\text{P}_J, m_{F}^{\prime}}$ by the 813.4 nm lattice light following the formalism in \cite{Dorscher_pra_2018}. We use the reduced matrix elements (RME) and transition frequencies for $^3\text{P}_1$ from \cite{Romaric2024} and derive the RME for $^3\text{P}_0$ transitions based on the line strengths and frequencies reported in \cite{Middelmann2014}.

Table \ref{Table: Raman rate} presents the calculated off-resonant scattering rates for a $\pi$-polarized optical lattice, where atoms initially in the $\ket{^3\text{P}_0,m_F=3/2}$ state (labeled as state $i$) scatter to $\ket{^3\text{P}_1,m_F}$ (state $j$). The subsequent decay rate to $\ket{^1\text{S}_0,m_F}$ is determined using Clebsch-Gordan coefficients. All rates are calculated for a trap depth of 1 $E{\text{rec}}$.

\begin{table}[!h]
\begin{center}

\begin{tabular}{||c c c c||} 
 \hline
 $i\rightarrow j$ & $F^{\prime}$ & $m_F^{\prime}$ & $\Gamma/(10^{-4}\text{s}^{-1})$ \\ [1 ex] 
 \hline\hline
  ${}^3\text{P}_0 \rightarrow{}^3\text{P}_1$ & 7/2 & 1/2 & 0.83 \\
  &  & 3/2 & 0. \\ 
  &  & 5/2 & 0.17 \\
  
  &9/2&1/2 & 1.21 \\ 
  &  & 3/2 & 0 \\ 
  &  & 5/2 & 1.06 \\

  &11/2&1/2 & 0.46 \\ 
  &  & 3/2 & 0 \\ 
  &  & 5/2 & 1.27 \\
 \hline\hline
 ${}^3\text{P}_0 \rightarrow {}^1\text{S}_0$ & 9/2 & -1/2 & 0.95 \\
 &  & 1/2 & 0.68 \\ 
 &  & 3/2 & 2.01 \\ 
 &  & 5/2 & 0.79 \\
 &  & 7/2 & 0.53 \\ 
  \hline\hline
 ${}^3\text{P}_0 \rightarrow{}^3\text{P}_2$ &  &  & 2.85 \\
  \hline\hline
 ${}^3\text{P}_0 \rightarrow{}^3\text{P}_0$ &  &  & 5.01 \\
 \hline
\end{tabular}
\end{center}
\caption{Theoretically calculated Raman scattering rates from $\ket{^3\text{P}_0, m_F=3/2}$ to various $\ket{^3\text{P}_1, m_F}$ states at a lattice potential of 1~$E{\text{rec}}$, along with the subsequent decay rates to different $\ket{^1\text{S}_0, m_F}$ states.}
\label{Table: Raman rate}
\end{table}

\section{Theoretical discussion on coherence enhancement}
\label{Coherence enhancement theory}
In this section, we briefly discuss the origin of the enhanced clock coherence. We first consider a simple three-level model: $\ket{c_g}$, $\ket{c_e}$, and $\ket{e}$. Here, $\ket{c_g}$ and $\ket{c_e}$ represent the ground and excited states of the clock transition (clock subspace), while $\ket{e}$ belongs to the disjoint subspace where population leaving the clock subspace is converted into detectable erasures.\par For simplicity, we assume a general loss channel that transfers population out of the clock excited state, with a finite probability of decaying either to the clock ground state or to the erasure subspace. This channel effectively models a combination of radiative decay, Raman scattering, and blackbody-induced scattering. The impact of such a channel during clock interrogation can be described by a quantum channel with Kraus operators:
\begin{align}
    K_0(t) &= \ket{c_g}\bra{c_g} + \sqrt{e^{-\Gamma t}}\ket{c_e}\bra{c_e} +\ket{e}\bra{e}\\
    K_1(t)&=\sqrt{(1-r_e) (1-e^{-\Gamma t})}\ket{c_g}\bra{c_e}\\
    K_e(t)&=\sqrt{r_e (1-e^{-\Gamma t})}\ket{e}\bra{c_e}\\
\end{align}

Where $\Gamma$ is the loss rate out of $\ket{c_e}$, and the fraction of loss that falls into the erasure subspace is denoted by $r_e$.  $K_0$ describes how the state evolves when no loss occurs. Specifically, the population in $\ket{c_g}$ and $\ket{e}$ remains the same while the probability of $\ket{c_e}$ damps. $K_1$ represents the probability of population scattered from $\ket{c_e}$ to $\ket{c_g}$, and $K_e$ represents the probability scattering to $\ket{e}$.\par
\begin{figure}[ht!]
\includegraphics[]{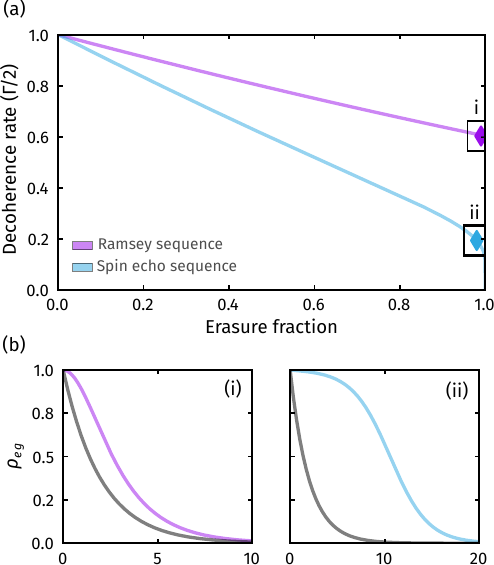}
\caption{\label{Fig_erasure_cal} (a) shows the theoretical decoherence rate for Ramsey (purple solid line) and spin echo (blue solid line) sequences with varying erasure fractions, assuming the only decoherence contribution is from erasure-convertible channels such as Raman scattering, loss, and radiative decay. (b) shows the calculated coherence versus time in the Ramsey sequence (i) with an erasure fraction of 1 and the spin echo sequence(ii) with an erasure fraction of 0.99. The black line in (i)\&(ii) represents the coherence without erasure conversion.}
\end{figure}

The standard readout method makes no distinction between the population in the clock subspace and the population in the erasure subspace. The evolution of the system corresponds to the full quantum channel and the coherence decays exponentially with a rate of $\Gamma/2$. However, the hyperfine-resolved readout gives access to the information solely in the clock subspace. Conditioning on not witnessing erasures, equivalent to only measuring population in the clock subspace, the system is subject to operations defined by $K_0$ and $K_1$. The outcome density matrix requires a proper renormalization as $K_0$ and $K_1$ together do not describe a complete quantum channel. The density matrix after evolution with hyperfine-resolved readout is in Eq.\ref{density_matrix_evolution} and the coherence is shown in Eq.\ref{coherence_with_erasure}
\begin{equation}
    \label{density_matrix_evolution}
    \rho = \frac{K_0\rho_0 K_0^{\dagger}+K_1\rho_0 K_1^{\dagger}}{\text{Tr}(K_0\rho_0 K_0^{\dagger}+K_1\rho_0 K_1^{\dagger})}
\end{equation}
\begin{equation}
    \label{coherence_with_erasure}
    \rho_{eg} = \frac{e^{-\frac{\Gamma}{2}t}/2}{1+(e^{-\Gamma t}-1)r_e/2}
\end{equation}
In Fig.\ref{Fig_erasure_cal}(a), we show the theoretically calculated coherence time as a function of the erasure fraction. The coherence time (defined as the $1/e$ time) increases with the erasure fraction, showing a $\times 1.6$ improvement at full erasure conversion. However, in this limit, the coherence time does not diverge to infinity, as $K_0$ is a non-Hermitian no-jump operator that induces wavefunction collapse upon a no-detection event\cite{Pradeep_PRL_2024}.

For the spin echo sequence, shown in Fig.\ref{Fig_erasure_cal}(c), the population in the clock subspace is inverted by a $\pi$-pulse, effectively redistributing the probability of decay back to the initial ground state. At high erasure fractions, the probability of correctly inferring the likely origin of the erasure is suppressed. As shown in Fig.\ref{Fig_erasure_cal}, the coherence time increases monotonically as the erasure fraction approaches 1, ultimately limited by dephasing sources rather than loss mechanisms.\par
In the ensemble clock experiment, the inclusion of two-body collision-induced loss and density shifts introduces additional complexities in the dependence of coherence time on the erasure fraction. The density shift is incorporated as a phenomenological dephasing term to match experimental data, though a rigorous treatment would require solving the many-body dynamics. 

\section{Numerical simulation of the experiment}
\label{Appendix: Numerical simulation}
In this section, we present the numerical model of the experiment, incorporating additional loss and dephasing mechanisms such as two-body collision loss of \tripletPzero and dephasing from elastic scattering. We account for an effective trap depth by considering the finite radial temperature, following the formulation in Ref.~\cite{Dorscher_pra_2018}. The system dynamics are modeled using the density matrix formalism. Since the experiment involves simultaneous clock interrogation, laser phase noise does not affect atom-atom coherence and is therefore omitted from the simulation.  In this framework, states 1, 2, and 3 correspond to the clock ground state, the clock excited state, and the remaining ground states, respectively.
\begin{align}
\begin{split}
    \dot{\rho}_{11}&=-\Gamma_{\text{L}}^{\text{g}}\rho_{11}+(\alpha\Gamma_{\text{r}}+\beta\Gamma_{0})\rho_{11}\\
    \dot{\rho}_{22}&=-(\Gamma_{\text{L}}^{\text{e}}+\Gamma_{0}+\Gamma_{\text{r}}+\Gamma_{\text{t}}\rho_{22})\rho_{22}\\
    \dot{\rho}_{33}&=-\Gamma_{\text{L}}^{\text{g}}\rho_{33}+(1-\alpha)\Gamma_{\text{r}}\rho_{22}+(1-\beta)\Gamma_{0}\rho_{22}\\
    \dot{\rho}_{12}&=-\frac{1}{2}(\Gamma_{\text{L}}^{\text{g}}+\alpha\Gamma_{\text{r}}+\beta\Gamma_{0}+\Gamma_{\text{t}}\rho_{22})\rho_{12}\\
    &\:\:\:\:\:-\Gamma^{e}_{\text{t}}(\rho_{11}+\rho_{22}+\rho_{33})\rho_{12}-i\Delta\rho_{12}\\
    \dot{\rho}_{21}&=-\frac{1}{2}(\Gamma_{\text{L}}^{\text{g}}+\alpha\Gamma_{\text{r}}+\beta\Gamma_{0}+\Gamma_{\text{t}}\rho_{22})\rho_{21}\\
    &\:\:\:\:\:-\Gamma^{e}_{\text{t}}(\rho_{11}+\rho_{22}+\rho_{33})\rho_{21} + i\Delta\rho_{21}\\
\end{split}
\end{align}
where $\Gamma_{\text{L}}^{\text{g}}$ is the one-body loss rate ($2.1\times10^{-2}$ s$^{-1}$) for ground-state atoms due to background gas collisions. $\Gamma_{\text{L}}^{\text{e}}$ is the one-body loss rate ($2.7\times10^{-2}$ s$^{-1}$$E_{\text{rec}}$$^{-1}$ + $1.8\times10^{-2}$ s$^{-1}$) for the excited state, accounting for both background gas collisions and Raman scattering to \tripletPtwo. $\Gamma_r$ is the Raman scattering rate to \tripletPone ($4.6\times10^{-2}$ s$^{-1}$$E_{\text{rec}}$$^{-1}$). $\Gamma_0$ is the natural lifetime of \tripletPzero (167 s).

The branching ratios of scattering and decay back to the clock subspace, $\alpha$ and $\beta$, are set to 0.18 and 0.42, respectively. $\Gamma_{\text{t}}$ is the two-body loss rate coefficient ($11\times10^{-6}$ s$^{-1}$cm$^{-3}$). $\Delta$ is the laser detuning on the clock transition. All rates above are extracted from independent measurements of the \tripletPzero natural lifetime~\cite{Jack_lifetime_2025}. Quenching from blackbody radiation (BBR) provides an additional, but minor, contribution that would affect the erasure fraction. However, the overall BBR-induced decay rate out of \tripletPzero is smaller than the natural radiative lifetime, and quenching via \tripletPtwo contributes only at second order. Consequently, BBR is not expected to be a significant source of error and is therefore neglected in the model.

We also incorporate a density-dependent phenomenological dephasing rate in the simulation. Notably, including a density-dependent dephasing term of $7.2\times10^{-12}$ s$^{-1}$cm$^{-3}$) is essential for achieving a good agreement between theory and experiment. We interpret this as capturing the effects of elastic scattering, similar to previous studies\cite{Bishof2011}, though a complete description of elastic collisions would require a full many-body treatment of the clock ensemble.

In the simulation, we fit the total number of atoms remaining in the lattice at the end of the experiment and use the fitted atom number at $t=0$ as the initial condition. When modeling the Ramsey experiment, we account for decoherence induced by magnetic field gradients within the ensemble. This is done by estimating the atomic distribution, assigning detunings based on axial positions, and averaging the coherence contribution across the ensemble.

Additionally, we incorporate a phenomenological Gaussian decay $T_2^*$ in the simulation to account for slow noise sources recoverable by the spin echo sequence, such as residual Stark shifts from the lattice. \par

The coherence $C(T)$ is calculated as the off-diagonal matrix element $|\rho_{12}(T)|^2$ in the density matrix. The population $N(T)$ is the sum of the diagonal matrix elements. In the case of hyperfine readout, only the population remaining in the clock subspace is counted, and $N(T)=|\rho_{11}(T)|^2+|\rho_{22}(T)|^2$. In the standard readout, the population sums over all diagonal matrix elements, including the remaining ground state population. Based on the simulated coherence $C(T)$ and atom number $N(T)$ as functions of interrogation time, we extract the QPN-limited Allan deviation using
\begin{equation}\label{appendix numerical simulation adev}
\sigma_{\text{QPN}}(T) = \frac{\sqrt{2}S(\phi,C(T))}{2 \pi \nu C(T) T} \sqrt{\frac{T+T_d}{N(T) \tau}}
\end{equation}
where $\nu$ is the clock transition frequency, $C(T)$ is the simulated atomic coherence, $T$ is the coherent interrogation time, and $T_d$ is the dead time between experimental shots, and $N(T)$ is the number of atoms.
For standard and reconstructed readout, $T_d = 1.75$ s, while for the hyperfine-resolved readout sequence, $T_d = 1.95$ s.\par
For differential clock measurements, the quantum projection noise limit leads to an additional scaling factor in the stability of the clock comparison that depends on the accumulated phase difference between the two ensembles as well as the contrast of the ellipse \cite{zheng_differential_2022, young_nature_2021,marti_2018}. We include this effect as an additional factor $S(\phi, C(T))$ given by
\begin{equation}\label{Appendix: numerical simulation scale factor}
\begin{split}
    &S  = \Big(\int_0^{2\pi}\frac{d\theta}{4 \pi}\frac{1}{\text{csc}^2(\theta)\Tilde{\sigma}^2(x)+\text{csc}^2(\theta+\phi)\Tilde{\sigma}^2(y)}\Big)^{-\frac{1}{2}}\\
    &\Tilde{\sigma}^2(x) = \frac{C(T)}{2}\text{cos}(\theta)-\frac{C(T)^2}{4}\text{cos}^2(\theta)\\
    &\Tilde{\sigma}^2(y) = \frac{C(T)}{2}\text{cos}(\theta+\phi)-\frac{C(T)^2}{4}\text{cos}^2(\theta+\phi)\\
\end{split}
\end{equation}
where $\theta$ represents the clock laser's phase and $\phi$ is the relative phase between the two ensembles. Because our differential clock measurement interrogates atoms beyond the atom-laser coherence time, the scaling factor is averaged over $\theta$ and thus depends on the relative phase $\phi$ and the contrast of the ellipse.
 For the Ramsey measurements, this relative phase evolves over time due to the detuning between the two ensembles, resulting in an oscillating and non-trivial dependence of stability with interrogation time $T$. We therefore only consider an idealized Allan deviation based on the simulated coherence and number of atoms without accounting for the additional effect of the phase, i.e., $\phi=0$, which represents a lower bound on the Ramsey stability. For the spin echo measurements, the accumulated phase difference is the result of the applied magnetic field gradient and is independent of interrogation time (as described in Appendix~\ref{Appendix_B}), so we calculate the expected stability accounting for the applied phase difference i.e. $\phi=0.04$ radians.\par

\section{Discussion of technical limitations}
\label{Appendix: Technical limitations}
As discussed in the main text, the two dominant error sources in the experiment are $\pi$-pulse infidelity and off-resonant scattering by the clock laser.\par

The $\pi$-pulse infidelities primarily arise from the phase noise of our local oscillator (LO) and atomic temperature variations. LO phase noise can be mitigated by operating at a higher Rabi frequency or by reducing laser noise through a longer cavity spacer with improved thermal shielding. Atomic temperature variations lead to inconsistent Rabi frequencies for atoms in different axial lattice bands, as well as radial inhomogeneity in regions of weaker confinement. Employing a larger lattice beam waist and deeper cooling is expected to mitigate these effects. By addressing these issues, we anticipate achieving a clock $\pi$-pulse fidelity of up to 99.5\%.

Additionally, off-resonant scattering of the clock light from adjacent $m_F$ states exacerbates the problem. At a bias magnetic field of approximately 5.5 G, the separation between adjacent $m_F$ states is around $\Delta \approx 2\pi \times 1000$ Hz, while the Rabi frequency is $\Omega \approx 2\pi \times 150$ Hz, leading to a few percent of non-resonant excitation. Since off-resonant scattering scales as $\Omega^2 / (\Omega^2 + \Delta^2)$, operating at a bias field of ~50 G and increasing $\Omega$ by a factor of 4 to improve the fidelity of the clock pulse could reduce off-resonant scattering by approximately a factor of 7.

\section{Coherence and Allan deviation at all trap depths}
\label{Appendix: additional plots}
\begin{figure*}[h!]
\includegraphics[]{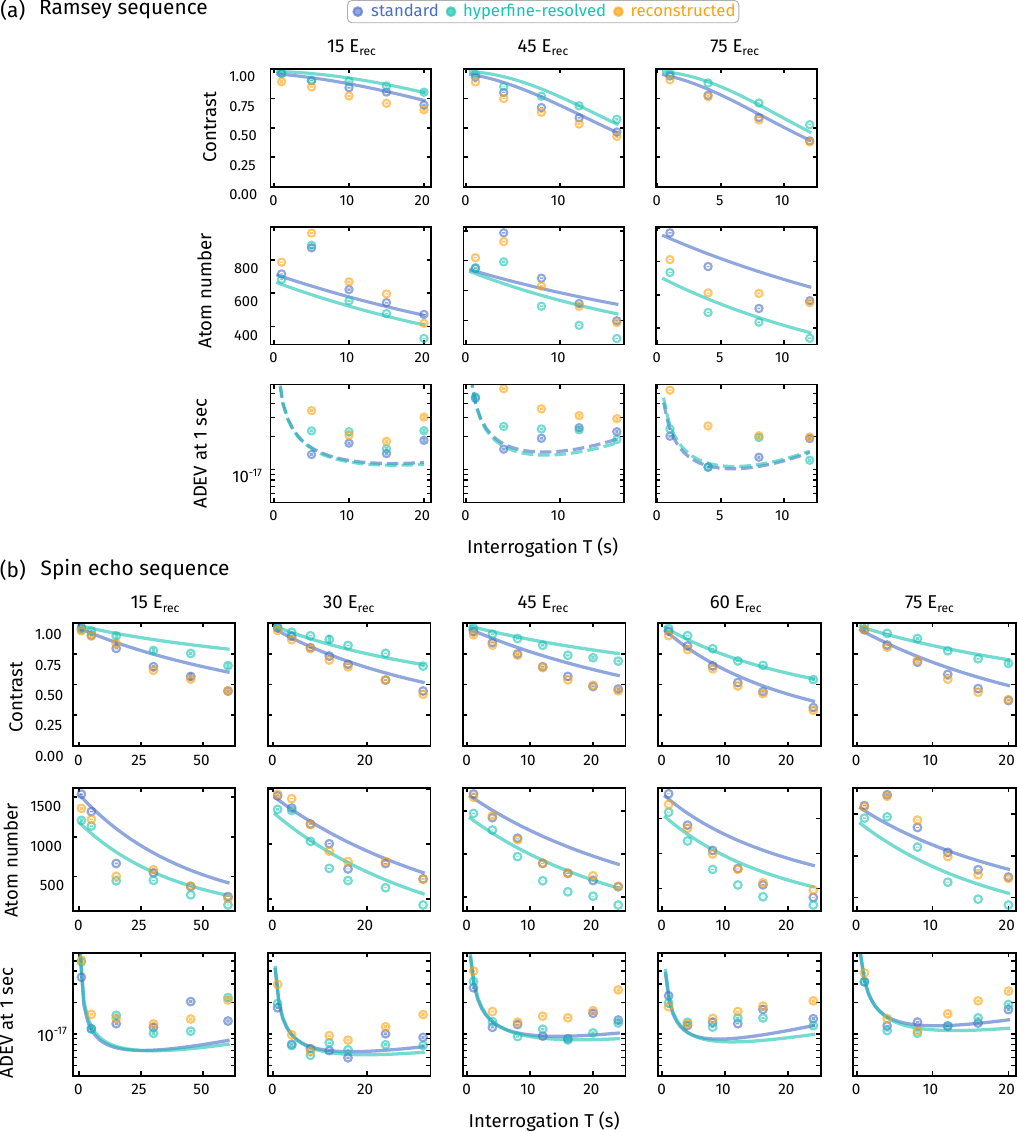}
\label{fig8_no_labels}
\caption{\label{Fig_8}Coherence, atom number and clock stability as functions of clock interrogation time at all trap depths measured (from 15 to 75 E\textsubscript{rec}) for (a) Ramsey sequence and (b) spin echo sequence. The simulated curves are plotted with the same color coding. The dashed curves in (a) represent the calculated stability using the simulated coherence and the remaining number of atoms, assuming the relative phase difference between ensembles is 0. The solid curves in (b) represent the calculated stability accounting for the effect of the relative accumulated phase between the two ensembles.}
\end{figure*}

In this section, we present the results of coherence and clock stability as functions of clock interrogation time in Fig.\ref{Fig_8}, measured at the trap depths described in the main text. The simulated coherence decay and calculated stability curves are also presented. The stability plot for the Ramsey sequence, shown in Fig.\ref{Fig_8} (a),  includes an idealized modeled stability (dashed curve) accounting for the simulated coherence and number of atoms calculated based on Eq.\ref{appendix numerical simulation adev}. In the spin echo sequence plot, shown in Fig.\ref{Fig_8} (b), we included a modeled stability (solid curves) that accounts for the relative accumulated phase between the two ensembles, as discussed in Appendix \ref{Appendix: Numerical simulation}.

\bibliography{erasure}

\end{document}